\documentclass[twocolumn,amsmath,amssymb,aps,prl,floatfix,nofootinbib,superscriptaddress]{revtex4}

\usepackage{graphicx,graphics,times,color}
\usepackage{bm}
\usepackage{longtable}

\def\be{\begin{equation}}
\def\ee{\end{equation}}
\def\ba{\begin{eqnarray}}
\def\ea{\end{eqnarray}}
\def\la{\langle}
\def\ra{\rangle}
\def\a{\alpha}

\begin{document}

\title{Entanglement Enhanced Information Transfer through Strongly Correlated Systems and its Application to Optical Lattices}
\author{Song Yang}
\affiliation{Department of Physics and Astronomy, University College
London, Gower St., London WC1E 6BT, United Kingdom}
\affiliation{Key Laboratory of Quantum Information, University of
Science and Technology of China, Hefei, 230026, People's Republic of
China}
\author{Abolfazl Bayat}
\affiliation{Department of Physics and Astronomy, University College
London, Gower St., London WC1E 6BT, United Kingdom}
\author{Sougato Bose}
\affiliation{Department of Physics and Astronomy, University College
London, Gower St., London WC1E 6BT, United Kingdom}

\date{\today}

\begin{abstract}
We show that the inherent entanglement of the ground state of strongly correlated systems can be exploited for both classical and quantum communications. Our strategy is based on a single qubit rotation which encodes information in the entangled nature of the ground state. In classical communication, our mechanism conveys more than one bit of information in each shot, just as dense coding does, without demanding long range entanglement. In our scheme for quantum communication, which may more appropriately be considered as a remote state preparation, the quality is higher than the highly studied attaching scenarios. Moreover, we propose to implement this new way of communication in optical lattices where all the requirements of our proposal have already been achieved.
\end{abstract}

\maketitle

{\em Introduction:--}
Strongly correlated systems often have nontrivial entangled states of multiple particles as their ground states.  When one wants to use the {\em dynamics}, as opposed to measurements  \cite{cluster-state}, of such a system for propagating information \cite{bose03}, the inherent entanglement plays little role and only symmetries of the state and the Hamiltonian seem to be important \cite{bayat-xxz}. The only mode of transmission studied so far is to attach a qubit encoding an ``unknown" quantum state to the system \cite{bose03,bayat-xxz}, which is motivated by the need to link quantum registers. This mode of transmission does not seek to harness the entanglement of many-body systems. Note that the entanglement in strongly correlated system is notoriously short ranged \cite{short-range} (with rare exceptions \cite{venuti}), so that it cannot be {\em directly} used to teleport \cite{Bennett-teleportation} an unknown state, accomplish remote state preparation \cite{bennett-remote} of a known state, or double the rate of classical communication by dense-coding \cite{Bennett-densecoding}. Thus, a natural question is whether the entanglement in many-body systems can practically benefit some mode information transfer. Many communication features, such as quantum key distribution, requires only ``known" states to be sent which may make the realization of the quantum communication simpler.
Thus, it is highly desirable to design new protocols for dynamical communication through strongly correlated systems, which are ``dense-coding-like" or ``remote state preparation-like" despite the {\em absence} of long range entanglement.

Cold atoms in an optical lattice is now an established field for many-body experiments. Both bosons \cite{Mott-insulator-boson} and fermions \cite{Mott-insulator-fermion} have been realized in the Mott insulator phase, where there is exactly one atom per site, and by properly controlling the intensity of laser beams one can get an effective spin Hamiltonian \cite{Lukin} between atoms.
Superlattices have been used to do singlet-triplet measurements of simulated spins in such systems \cite{singlet-triplet,Trotzky-PRL,Rey-PRL}.
Striking new developments \cite{Medley2010,single-site,Meschede,Weitenberg,Gibbons-Nondestructive} of 2010/11 make it timely to seriously consider the implementation of communication schemes in optical lattices.
New cooling techniques  \cite{Medley2010} have enabled, reaching for the first time, the temperatures required for observing quantum magnetic phases.
Moreover, single atom detection \cite{single-site} with single-site resolution as well as single qubit operation and measurement
\cite{Meschede,Weitenberg,Gibbons-Nondestructive}  have been achieved. In a nutshell, operating on individual atoms in an optical lattice is rapidly becoming a viable procedure. In view of these, near-future experiments are likely to look at the effects of local actions, e.g. spin flips \cite{Meschede,Weitenberg}, in optical lattice simulated spin systems, primarily to look at propagation of spin waves from a condensed matter angle. A good question then is whether the {\em same} experiments can also provide interesting quantum information protocols.

In this letter, we introduce a new mechanism for both quantum and classical communications in spin chains which uses a local rotation by a sender, followed by non-equilibrium dynamics, and subsequent reception and measurements by a receiver. Surprisingly, despite the {\em absence} of long range entanglement we find that even a single qubit rotation can convey {\em more} than one bit of information in the same spirit used by the dense coding protocol \cite{Bennett-densecoding}. We also find that the fidelity of quantum state transfer is enhanced compared to the fidelity achieved by attaching a qubit to the system.

\begin{figure} \centering
    \includegraphics[width=6cm,height=4cm,angle=0]{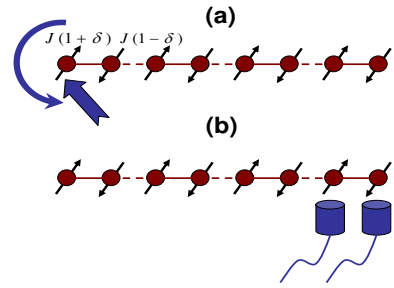}
    \caption{(Color online) (a) A dimerized chain where solid (dashed) lines represent strong (weak) bonds and encoding is achieved through a local rotation on qubit 1. (b) Decoding is done through measurements on qubits $N-1$ and $N$.}
     \label{fig0}
\end{figure}

{\em Setup:-} We consider a chain of $N$ spin-$1/2$ particles, where $N$ is even, interacting through a dimerized Hamiltonian
\begin{equation}
\label{eq:H}
H=J\sum\limits_{k=1}^{N-1}(1+(-1)^{k+1}\delta) \hat{\sigma}_{k}\cdot\hat{\sigma}_{k+1},
\end{equation}
where, $J>0$ is the coupling, $\hat{\sigma}_{k}=(\sigma_k^x,\sigma_k^y,\sigma_k^z)$ denotes the Pauli operators at site $k$ and $0<\delta<1$ determines the dimerization of the chain. A schematic of this system has been shown in Fig. \ref{fig0}(a).
We assume that system is initially in its ground state $|GS\ra$. Due to the $\text{SU(2)}$ symmetry of the Hamiltonian, the reduced density matrix of the first two qubits is a Werner state $\rho_{_{1,2}}= p|\psi^{-}\rangle\langle\psi^{-}|+(1-p)\hat{I}_4/4$, where $\hat{I}_n$ represents an $n\times n$ identity matrix, $|\psi^{-}\rangle=(|01\rangle-|10\rangle)/\sqrt{2}$ is the singlet state and, $0\leq p\leq 1$ is controlled by $\delta$ (if $\delta\rightarrow 1$ then $p\rightarrow 1$). We assume that sender (Alice) controls qubit 1 as shown in Fig. \ref{fig0}(a) while receiver (Bob) controls the qubits $N-1$ and $N$ as shown in Fig. \ref{fig0}(b).

\begin{figure} \centering
    \includegraphics[width=9cm,height=4cm,angle=0]{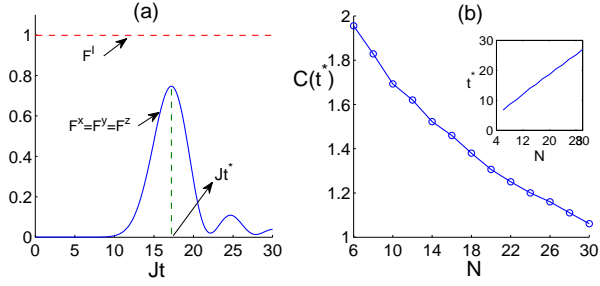}
    \caption{(Color online) (a) $F^\a$ as a function of time for a chain of $N=20$ and $\delta=0.7$. (b) Scaling of $C(t^*)$ in terms of $N$. Inset shows the scaling of the optimal time $t^*$ versus $N$ ($\delta=0.7$).}
     \label{fig1}
\end{figure}

{\em Classical communication:--} For classical communication,
Alice encodes two bits of classical information (i.e., $00$, $01$, $10$ and $11$) in the state of the chain by a single operation
$\sigma_{1}^\alpha$ on qubit 1, where $\alpha \in \{I, x, y, z, \}$ and $\sigma_{1}^I=\hat{I}_2$
such that $\sigma_1^I$ represents $00$, $\sigma_1^x$ represents $01$ and etc.
Unlike original dense coding proposal \cite{Bennett-densecoding}, this encoding is local since there is no long-range entanglement between Alice and Bob. Accordingly, the quantum state of the whole system changes to $|\psi^\a(0)\ra=\sigma_1^\alpha|GS\ra$.
The reduced density matrix of the first two qubits after this local action becomes $\rho_{_{1,2}}^\a(0)=\sigma_{1}^\alpha\rho_{_{1,2}}\sigma_{1}^\alpha=
p|b^\a\rangle\langle b^\a|+(1-p)\hat{I}_4/4$ in which the singlet part of the $\rho_{_{1,2}}$ has been replaced by one of the four Bell states $|b^\alpha\ra=\sigma_{1}^\alpha|\psi^-\rangle$.
Notice that unlike the ideal case (where $p=1$), for $p<1$ the encoded states are not fully distinguishable as they are all mixed with identity.

After encoding, system evolves as $|\psi^\a(t)\ra=e^{-iHt}|\psi^\a(0)\ra$ and time evolution of the system transfers $\rho_{_{1,2}}^\a(0)$ dispersively along the chain. At time $t$ the density matrix of qubits $N-1$ and $N$ is $\rho_{_{N-1,N}}^\a(t)$ for which Bob can define a super operator $\mathcal{E}$ as $\rho_{_{N-1,N}}^\a(t)=\mathcal{E} (\rho_{_{1,2}}^\a(0))$. At an optimal time $t=t^*$, the density matrix $\rho_{_{N-1,N}}^\a(t^*)$ has its maximal fidelity with $\rho_{_{1,2}}^\a(0)$ and by performing Bell measurement on qubits $N-1$ and $N$ Bob can identify the operator $\sigma_1^\a$ (and accordingly two classical bits encoded by Alice) through his measurement outcome $|b^\alpha\ra$. However, Bob may have some errors in his decoding as: (i) the initial encoding may not be perfect ($p<1$); (ii) dynamics is dispersive and the received state $\rho_{_{N-1,N}}^\a(t^*)$ is not exactly equal to $\rho_{_{1,2}}^\a(0)$.
To quantify the quality of communication for each $\sigma_1^\alpha$ one can numerically compute the fidelity $F^\alpha(t) = \langle b^{\alpha}|\rho_{_{N-1,N}}^\a(t)|b^{\alpha}\rangle$. In Fig.~\ref{fig1}(a) we plot $F^\a(t)$ versus time for a chain of $N = 20$ in which fidelity peaks at the time $t=t^*$.

In classical communication, Holevo information is usually employed to get a quantification of the amount of information which is sent. For our proposed mechanism, the Holevo information is
$C(t) = S(\mathcal{E}(\sum_{\a} q_{a} \rho_{_{1,2}}^\a))-\sum_{\a} q_{\a}S(\mathcal{E}(\rho_{_{1,2}}^\a))$, where, $S(\rho)=-tr(\rho \log \rho)$ is the von Neumann entropy and $q_\a$ is the probability of applying $\sigma_1^\a$. We assume equiprobable inputs, i.e., $q_\a=1/4$, which also maximizes $C(t)$.
Holevo information also peaks at $t=t^*$ and with time dependent Density Matrix Renormalization Group (tDMRG) techniques we numerically simulate systems up to 30 sites.
In Fig.~\ref{fig1}(b) we plot $C(t^*)$ as a function of $N$.
It shows that $C(t^*)$ is still above 1 bit for a chain of length $N=30$ (for $\delta=0.7$). In the inset of
Fig.~\ref{fig1}(b) we plot the optimal time $t^*$ versus $N$ which linearly increases with length.

\begin{figure} \centering
    \includegraphics[width=9cm,height=4cm,angle=0]{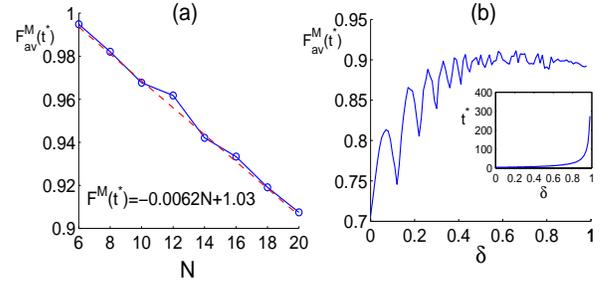}
    \caption{(Color online) $F_{av}^M(t^*)$ as a function of $N$ (blue circles) and its linear fit (red dashed line) for a chain with $\delta=0.7$. (b) $F_{av}^M(t^*)$ as a function of dimerization $\delta$ for a chain of $N=20$. Inset shows optimal time $t^*$ as a function of $\delta$. }
     \label{fig2}
\end{figure}

{\em Quantum communication:--} We can use the same recipe for quantum communication and remote state preparation. To encode one qubit in a pure state Alice applies
\begin{equation}
\label{eq:R1}
R_1(\theta,\phi)=
\begin{pmatrix}
\cos \frac{\theta}{2} & -\sin \frac{\theta}{2} e^{-i\phi}\\
\sin \frac{\theta}{2} e^{i\phi}& \cos \frac{\theta}{2}
\end{pmatrix},
\end{equation}
on the first qubit of the chain. The application of the operator $R_1$ on state $|0\ra$ (or $|1\ra$) gives the most general pure state of a qubit on the surface of the Bloch sphere determined by two angles $\theta$ and $\phi$. After the operation of $R_1$, state of the system changes to $|\psi^{\theta,\phi}(0)\ra=R_1(\theta,\phi)|GS\ra$ and the reduced density matrix of the first two qubits become
$\rho_{_{1,2}}^{\theta,\phi}(0)=pR_1|\psi^-\ra \la \psi^-|R_1^\dagger+(1-p)\hat{I}_4/4$. After encoding, system evolves as $|\psi^{\theta,\phi}(t)\ra=e^{-iHt}|\psi^{\theta,\phi}(0)\ra$ and the density matrix of Bob's qubits, i.e. $\rho_{_{N-1,N}}^{\theta,\phi}(t)$, accordingly changes with time.
At time $t=t^*$, the two parameters of $R_1$ (i.e. $\theta$ and $\phi$) are encoded in the density matrix of Bob's qubits, i.e. $\rho_{_{N-1,N}}^{\theta,\phi}(t^*)$ and he can localizes this information in a single qubit by performing a single-qubit measurement in the computational basis on site $N-1$. In an ideal case, where both encoding and transmission are perfect, Bob receives $R_1|\psi^-\ra \la \psi^-|R_1^\dagger$. One can easily show that in the state $R_1|\psi^-\ra$ when qubit $N-1$ is projected on $|0\ra$ or $|1\ra$ the state of qubit $N$ is collapsed according to
\begin{eqnarray}
\label{eq:Measurement}
    |0\rangle_{_{N-1}}&\rightarrow& |\psi_0\ra_{_N}=\cos\frac{\theta}{2}|1\rangle+\sin\frac{\theta}{2}e^{-i\phi}|0\rangle \cr
    |1\rangle_{_{N-1}}&\rightarrow& |\psi_1\ra_{_N}=\cos\frac{\theta}{2}|0\rangle-\sin\frac{\theta}{2}e^{+i\phi}|1\rangle,
\end{eqnarray}
where, $|\psi_k\ra_N$ ($k=0,1$) is the state of site $N$ when site $N-1$ is projected in state $|k\ra$.
However, in a realistic situation qubit $N$ remains mixed even after measuring qubit $N-1$. So, the measurement fidelity is defined as
$F^M(\theta,\phi,t)=p_0\la 0, \psi_0|\rho_{_{N-1,N}}^{\theta,\phi}(t)|0, \psi_0\ra+ p_1\la 1, \psi_1| \rho_{_{N-1,N}}^{\theta,\phi}(t) |1, \psi_1\ra$ where, $p_k$ denotes the probability of projecting qubit $N-1$ on state $|k\ra$. To have an input independent quantity, we compute the average measurement fidelity by integrating $F^M(\theta,\phi,t)$ over the surface of the Bloch sphere as $F_{av}^M(t)=\int{F^M(\theta,\phi,t) d\Omega}$. With considering the exact form of $R_1$ in Eq. (\ref{eq:R1}) one can analytically show that
\begin{eqnarray}
F_{av}^{M}(t) &=& \frac{1}{2}+\frac{1}{12}(F_2 - F_1)+\frac{2}{3} F_3,
\end{eqnarray}
where,
\begin{eqnarray}
F_1 &=& \langle GS|\sigma^{z}_{N-1}(0)\sigma^{z}_{N}(0)|GS \rangle,\cr
F_2 &=& 2 \langle GS|\sigma^{+}_{1}(0)\sigma^{z}_{N-1}(t)\sigma^{z}_{N}(t)\sigma^{-}_{1}(0) |GS \rangle,\cr
F_3 &=&  \mathrm{Re}(\langle GS|\sigma^{+}_{1}(0)\sigma^{z}_{N-1}(t)\sigma^{-}_{N}(t)|GS\rangle).
\end{eqnarray}
These components can be computed numerically with the means of exact diagonalization.
In Fig.~\ref{fig2}(a) we plot $F_{av}^M(t^*)$ as a function of $N$ for $\delta=0.7$. According to Fig.~\ref{fig2}(a), the average fidelity decays very slowly and fits by the line $F_{av}^M(t^*)=-0.0062N+1.03$. Remarkably, extrapolation shows that the average fidelity is above the classical threshold 2/3 for chains up to $N=58$.

Our protocol is fundamentally different from the usual attaching schemes because of the direct role of entanglement in our proposal. In fact, exploiting the inherent short range entanglement in strongly correlated systems improves the transmission quality.  We compare the average fidelities achieved in our proposal and ferromagnetic/anti-ferromagnetic attaching scenarios in TABLE I. As the numbers evidently show $F_{av}^M(t^*)$ is always higher than attaching schemes for both ferromagnetic and anti-ferromagnetic chains and improvement becomes even more significant in longer chains.

{\em Mechanism:-} In almost all works in the context of quantum communication through many-body systems (for instance see \cite{Hengl}) people first establish long range entanglement between the sender and receiver through entanglement propagation and then use that entanglement for teleportation. Our mechanism is fundamentally different as distributing entanglement between the sender and receiver is {\em not} our aim. We instead exploit the inherent local entanglement (i.e., between proximal spins) in the initial state of the system for communication. In the absence of local entanglement, for instance in a ferromagnetic initial state, the actions of $\sigma_1^x$ and $\sigma_1^y$ are identical (spin flip) and hence {\em cannot} be used for encoding different states. The capability of using localized entanglement for some tasks which in general need long range entanglement is the {\em unique} feature of strongly correlated systems that we point out.
To have a purer local entanglement, and so a better encoding, one has to use a proper $\delta$. In a chain of length $N=20$ for $\delta > 0.5$ we have $p>0.99$. On the other hand by increasing $\delta$ the propagation becomes slower due to the emergence of small couplings (i.e. $J(1-\delta)$) which favors intermediate values of $\delta$. In Fig.~\ref{fig2}(b) we plot $F_{av}^M(t^*)$ as a function of $\delta$ for a chain of length $N=20$ which goes up oscillatory to take its maximum value around $\delta=0.7$. In the inset of Fig.~\ref{fig2}(b) we plot $t^*$ as a function of $\delta$ which exponentially grows for $\delta > 0.8$.
For relatively large $\delta$, the ground state of the system is almost a series of singlets --- so instead of the $|GS \rangle$ a series of singlets can be prepared, which may be possible in an independent process which avoids sophisticated cooling. By starting from series of singlets, and a Hamiltonian with $\delta\geq 0.7$ for evolving the system, the figures of merit reached in both quantum and classical communications are almost the same as those for the $|GS \rangle$ as the initial state.

\begin{table}
\begin{centering}
\begin{tabular}{|c|c|c|c|c|c|c|c|c|}
  \hline
  $N$      &6         & 8     & 10    & 12    & 14    & 16    & 18    & 20 \\
  \hline
  FM   & 0.820 &0.787 &0.763 &0.745 &0.731 &0.719 &0.708 &0.699 \\
  \hline
  AFM  & 0.954 &0.935 &0.919 &0.906 &0.895 &0.885 &0.877 &0.871 \\
  \hline
  $F_{av}^M(t^*)$ & 0.993 &0.980 &0.967 &0.961 &0.941 &0.932 &0.918 &0.906\\
  \hline
\end{tabular}
\caption{Comparison between different strategies of quantum communication namely, anti-ferromagnetic (AFM) and ferromagnetic (FM) chains with attaching an extra spin as well as $F_{av}^M(t^*)$ achieved in our scheme.}
\par\end{centering}
\centering{}\label{table_1}
\end{table}

\begin{figure} \centering
    \includegraphics[width=7.5cm,height=5.5cm,angle=0]{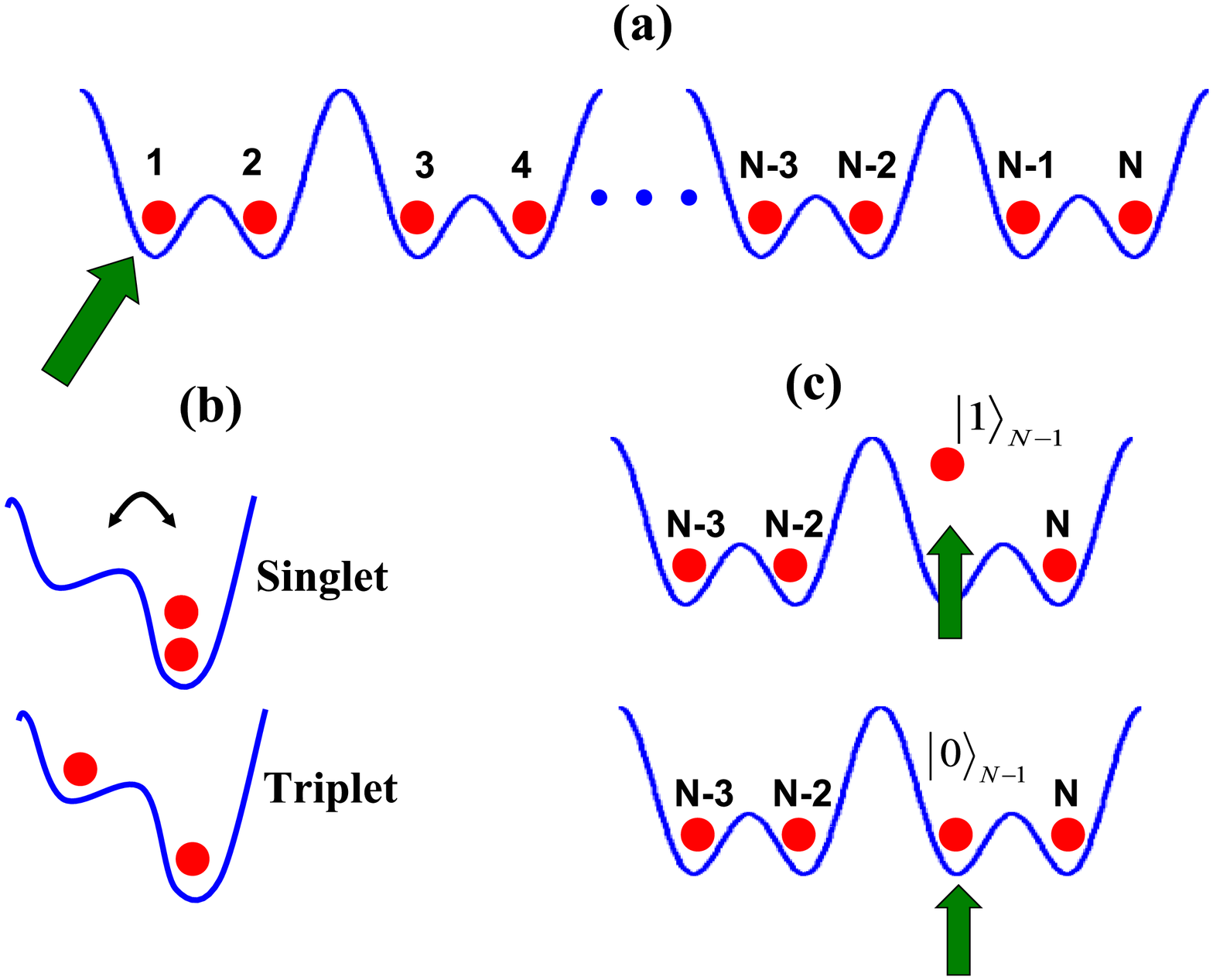}
    \caption{(Color online) Schematic figure for optical lattice realization. (a) Encoding through a local rotation on the first qubit. (b) Bell measurement through applying a state-dependent offset in which two atoms hop to a single site when their internal state is singlet and remain apart otherwise. (c) Single site measurement by an intense laser beam which pushes the atom of the lattice if it is in state $|1\ra$ and leaves it there if it is in $|0\ra$. }
     \label{fig3}
\end{figure}

{\em Application:--}
As an application for the above mechanism, we propose an array of cold atoms in their Mott insulator phase sitting in the minimums of a superlattice potential \cite{singlet-triplet,Trotzky-PRL,Rey-PRL} formed by counter propagating laser beams with two frequencies, which one being twice the other (Fig.~\ref{fig3}(a)). In the limit of high on-site energy the interaction between atoms is effectively modeled by a spin Hamiltonian \cite{Lukin}.
The alternating barriers between the atoms in a superlattice allows for realizing the dimerized Hamiltonian of Eq. (\ref{eq:H}).
For encoding information, classical or quantum, we need a unitary operation acting on site $1$ which is achieved by shining a laser on qubit $1$ as shown in Fig.~\ref{fig3}(a).
To have a local gate operation without affecting the neighboring qubits one may apply a weak magnetic field gradient \cite{Meschede}, or use a tightly focused laser beam \cite{Weitenberg} to split the hyperfine levels of the target atom. Then a microwave pulse, tuned only for the target qubit, operates the gate locally as has been realized in Refs.~\cite{Meschede,Weitenberg}.

When encoding finishes, Bob should wait for a period of $t^*$ and then decode information. As discussed before, in classical communication, decoding is a Bell measurement on sites $N-1$ and $N$ while, in quantum communication, it is a single site measurement on site $N-1$. Since the measurement itself takes time and in particular Bell measurement may not be very fast we may stop the dynamics at $t=t^*$ with raising the barriers quickly such that $J\rightarrow 0$.
Notice that in a superlattice one can control the even (odd) couplings independently with tuning the intensity of the low (high) frequency trapping laser beam \cite{singlet-triplet,Trotzky-PRL}.
Freezing the dynamics allows for slow measurements to be accomplished. For Bell measurement, used in classical communication, we can measure singlet-fraction through spin triplet blockade technique which has been recently proposed \cite{Rey-PRL} and then realized \cite{singlet-triplet,Trotzky-PRL}. In that method a spin dependent offset is applied on both sites to deform the superlattice such that one atom hops to the other site to have both atoms together only if their internal state is a singlet \cite{Rey-PRL} (Fig.~\ref{fig3}(b)). A subsequent fluorescent photography \cite{single-site}, which can be done without disturbing the internal states \cite{Gibbons-Nondestructive}, reveals the position of the atoms and therefore determines the singlet/triplet state of the atoms, however, it does not discriminate the non-singlet Bell states. To distinguish between non-singlet Bell states Bob has to apply a local Pauli rotation on qubit $N-1$ to convert one of the non-singlet Bell states to singlet and then a subsequent singlet/triplet measurement determines whether the new state is singlet or not. In the worst case with two local operations followed by singlet/triplet measurements, Bob accomplishes his Bell measurement on qubits $N-1$ and $N$.
On the other hand, in quantum communication in which single site measurement is expected on site $N-1$, we can use the technique of Refs.~\cite{Meschede}. In that methodology state $|1\ra$ is coupled to an excited state through an intense perpendicular laser beam whose radiation pressure pushes the atom out of the lattice. This leaves the site empty if its atom is in state $|1\ra$ and full if the atom is in state $|0\ra$ as shown schematically in Fig.~\ref{fig3}(c). This can thereby be read by fluorescent imaging.

{\em Initialization:-} Interestingly, {\em one does not need sophisticated cooling methods} to create the initial state of our protocol. An ideal initialization is a series of singlets ($\delta=1$) as long as the subsequent non-equilibrium dynamics happens at $\delta\neq 1$. This initial state has already been prepared \cite{singlet-triplet}. Independently, by starting from a band insulator and adiabatically changing the lattice potential as proposed in a very recent paper \cite{Cirac-AFM}, ground states with other $\delta$ can be realized as initial states. Thus, temperature only needs to be less than the band gap of the insulator ($\gg J$). A more direct strategy of cooling to the ground state requires the temperature of the system to be less than the energy gap ($\sim 4J(1+\delta)$). For a typical value of $J=360$ Hz \cite{Trotzky-PRL}, one finds $K_BT<100$~nK. Reaching this range of temperatures is at the edge of the current experiments \cite{temperature-lattice,Medley2010} and will soon be available.

{\em Conclusion:--} We introduced a new methodology for truly exploiting the inherent entanglement in the ground state of strongly correlated systems for both classical and quantum communications. In our proposed scheme a local rotation on a single qubit encodes information in the entangled ground state of the many-body system. In classical communication, this encoding enables conveying more than one bit of information, just as in dense coding, without any prior shared entanglement. We also showed that the same recipe can be used for quantum communication which gives a better quality in comparison to the usual attaching scenarios. Moreover, this proposal is especially timely in the context of optical lattice implementations where all the requirements of our proposal have been achieved in recent experiments. It will provide a quantum information angle to the foreseeable study of propagations of local spin flips \cite{Meschede,Weitenberg} in optical lattices.

{\em Acknowledgments.}
Discussion with A. Beige,
K. Bongs, D. Jaksch, D. Lucas, F. Renzoni and M. Lubasch
are kindly acknowledged.
SY thanks CSC for financial support.
AB and SB acknowledge the EPSRC. SB also thanks
the Royal Society and the Wolfson Foundation.

\end{document}